\documentstyle[12pt,epsf,epsfig]{article}

\font\grande=cmr10 scaled \magstep4
\font\medio=cmr10 scaled \magstep2
\outer\def\beginsection#1\par{\medbreak\bigskip
      \message{#1}\leftline{\bf#1}\nobreak\medskip
\vskip-\parskip
      \noindent}

\def\laq{\raise 0.4ex\hbox{$<$}\kern -0.8em\lower 0.62
ex\hbox{$\sim$}}
\def\gaq{\raise 0.4ex\hbox{$>$}\kern -0.7em\lower 0.62
ex\hbox{$\sim$}}
\def\beq{\begin{equation}}
\def\eeq{\end{equation}}
\def\bea{\begin{eqnarray}}
\def\eea{\end{eqnarray}}
\def\bean{\begin{eqnarray*}}
\def\eean{\end{eqnarray*}}

\begin{document}
\bibliographystyle {unsrt}

\titlepage
\begin{flushright}
CERN-TH/99-32 \\
hep-th/9902126 \\

\end{flushright}
\vspace{15mm}
\begin{center}
{\grande Pre-bangian origin of our entropy and time arrow}\\

\vspace{15mm}

     G. Veneziano \\
\vspace{6mm}

{\sl Theory Division, CERN, CH-1211 Geneva 23, Switzerland} \\

\end{center}

\vskip 2cm
\centerline{\medio  Abstract}

\noindent

I argue that, in the  chaotic version of string cosmology proposed recently,
classical and quantum effects generate, at the time of
exit to radiation, the correct amount of entropy to saturate
a Hubble (or holography) entropy  bound (HEB) and  to identify, within our own Universe, the arrow of time. Demanding that the HEB be fulfilled at all times forces a crucial ``branch change" to occur,
and the so-called string phase to end at a critical value of the
effective Planck mass, in agreement with previous conjectures.

\vspace{5mm}

\vfill
\begin{flushleft}
CERN-TH/99-32 \\
February 1999\\

\end{flushleft}

\newpage

The origin of the present entropy of our Universe, $S_0$, is one of the deepest cosmological
mysteries. The $2.7$ K cosmic microwave background (CMB), if it indeed fills
our observable Universe  uniformly, contributes a gigantic $10^{90}$ to $S_0$. However,
as repeatedly emphasized by many people, most notably by Roger Penrose \cite{Penrose},
 such an amount
falls very short of what entropy could have been expected  to be, even if we go back
to the Planckian era, i.e. to $t=t_P \sim 10^{-43}$ s after the big bang.
 Since entropy can only grow, the entropy of our Universe  at $t=t_P$,
$S_P$, must be smaller than $S_0$; yet, on the basis of the energy content
  and of the size of the Universe $R_P$ at $t\sim t_P \equiv l_P/c$, we might have expected \footnote{Throughout
this paper we will stress functional dependences
while ignoring numerical factors.}
\begin{equation}
 S_P \sim E_P R_P/c\hbar \sim \rho_P R_P^4 \sim (R_P/l_P)^4 \sim 10^{120} ~~.
\label{BB}
\end{equation}

The fact that the entropy of our Universe must have been at least $30$ orders
of magnitude smaller than the value in (\ref{BB})  would nicely  ``explain" our arrow of time,
 by identifying the beginning of the Universe with this state of incredibly small
entropy near the Planck time \cite{Penrose}. In order to solve the problem, Penrose \cite{Penrose} invokes, without
much justification, a new ``Weyl-curvature hypothesis".
The expected value given in Eq. (\ref{BB}) coincides with the so-called
Bekenstein entropy bound (BEB) \cite{BB}, which states that, for any physical system of energy $E$
and physical size $R$, entropy cannot exceed $S_{BB} = ER/c\hbar$. This bound is saturated by
a black hole of mass $E$ and size equal to its 
Schwarzschild radius $R=GE$. If the newly born Universe were a single black hole its Schwarzschild radius would have been
much larger than $R_P$, and an even higher
entropy, $O(10^{180})$, would have resulted. What could have made the
initial entropy much smaller than  $S_{BB}$ is instead
 the possibility that the Universe,
right after the big bang, was already in a very ordered, homogeneous state.
 But this is just restating the puzzle in terms of the usual homogeneity
problem of standard (non-inflationary) cosmology \cite{inflation}.

The way the two problems are related can be made explicit by introducing a stronger
bound on entropy, which, unlike Bekenstein's general bound, should apply to the special case of (fairly) homogeneous
cosmological situations. We shall call it the ``Hubble entropy bound"  and formulate it as follows:
Consider a  sufficiently homogeneous Universe in which a (local) Hubble expansion (or contraction) rate can be defined, in the sinchronous gauge, as:
\begin{equation}
 H \sim  1/6~\partial_t (\log g) ~~ , ~~ g \equiv {\rm det}~(g_{ij}) ~~,
\label{defH}
\end{equation}
with $H$ varying little (percentage-wise) over distances $O(H^{-1})$. In this case $H^{-1}$,
the so-called Hubble radius,
is known to correspond to the scale of causal connection, i.e. to the scale within which
microphysics can act. In such a context it is hard to imagine that a black hole
larger than $H^{-1}$ can  form, since, otherwise, different parts of its horizon would
be unable to  hold together. Thus, the largest entropy we may conceive is the one
 corresponding to having just one black hole per Hubble volume $H^{-3}$. Using
the Bekenstein--Hawking formula for the entropy of a black hole leads to our
proposal of a  ``Hubble entropy bound" (HEB):
\begin{equation}
S < S_{HB} \equiv l_P^{-2} \sum_{i} H_i^{-2} \sim n_H S_H ~~,
\label{HB}
\end{equation}
where the sum runs over each Hubble-size region. The last estimate in (\ref{HB}) assumes
a fairly constant  $H$ throughout space, and defines $n_H$ as the number of Hubble-size
regions, each one carrying maximal entropy $S_H = l_P^{-2}  H^{-2}$.
This bound appears to be related, at least in some cases, to the one recently proposed
 by Fischler and Susskind \cite{FS} on the basis of the so-called holography principle.
We may thus take $S_{HB}$ to stand for Hubble or Holography entropy bound according to taste.
If one applies the HEB to the initial Universe, one finds \cite{FS} that,
unlike the BEB, it is practically saturated. It is widely fulfilled thereafter \cite{FS}.

This letter aims at  explaining why saturation of the HEB naturally
 takes place at the big bang in the context of pre-big bang (PBB) cosmology \cite{PBB}.
Before proceeding, we note that, in ordinary inflation, the entropy problem is solved \cite{inflation} by invoking a
non-adiabatic reheating process occurring after inflation. Since inflation has
already made the Universe homogeneous, after thermal equilibrium is reached
entropy is given by
its standard thermodynamic relation to temperature (here the reheating temperature) as
\begin{equation}
S_{RH} \sim T_{RH}^3 R_{RH}^3 \sim S_0 ~ .
\label{SSI}
\end{equation}
One thus  naturally obtains the correct value. However,
unlike what will be shown to be the case in the PBB scenario,
 $S_{RH}$ fails to saturate the HEB since:
\begin{equation}
S_{HB}(t=t_{rh}) = l_P^{-2} H_{RH} R_{RH}^3 = H_{RH}^{-1} T_{RH}^4 R_{RH}^3 =
 (T_{RH}/ H_{RH}) ~ S_{RH} \gg S_{RH}\;.
\label{SIHB}
\end{equation}

In order to discuss various forms of entropy in the PBB scenario, let us recall some basic ideas,
which have emerged from recent studies of the latter (see \cite{Erice} for a review).
It now looks quite certain that generic --though  sufficiently weak-- initial conditions
lead to a form of stochastic PBB, which, in the Einstein-frame metric, can be seen as a
 sort of chaotic gravitational collapse \cite{GV,BDV}. Black holes of different
 sizes form but, for an observer
inside each horizon measuring distances with a stringy meter \footnote {Note that, while we shall  work in the string frame throughout, the  same results would also follow in the Einstein frame.}, this is experienced as
 a pre-big bang inflationary cosmology in which
the $t=0$ (hopefully fake) big bang singularity is identified \cite{BDV} with the (hopefully equally fake)
black hole singularity at $r=0$.

We are thus led to identifying our observable Universe as what became of a portion of space
that was originally inside a large enough black hole.
In general, if we want to achieve a very flat and homogeneous Universe, we should better identify our present Universe with just a tiny piece of the
collapsing/inflating region. For the purpose of this note, however, this would only complicate
the equations without adding new physical information. This is why, hereafter, we shall
identify our present Universe with the  whole interior of a single initial black hole.

\newpage

\begin{figure}
\hglue 1.6 cm
 \epsfig{figure=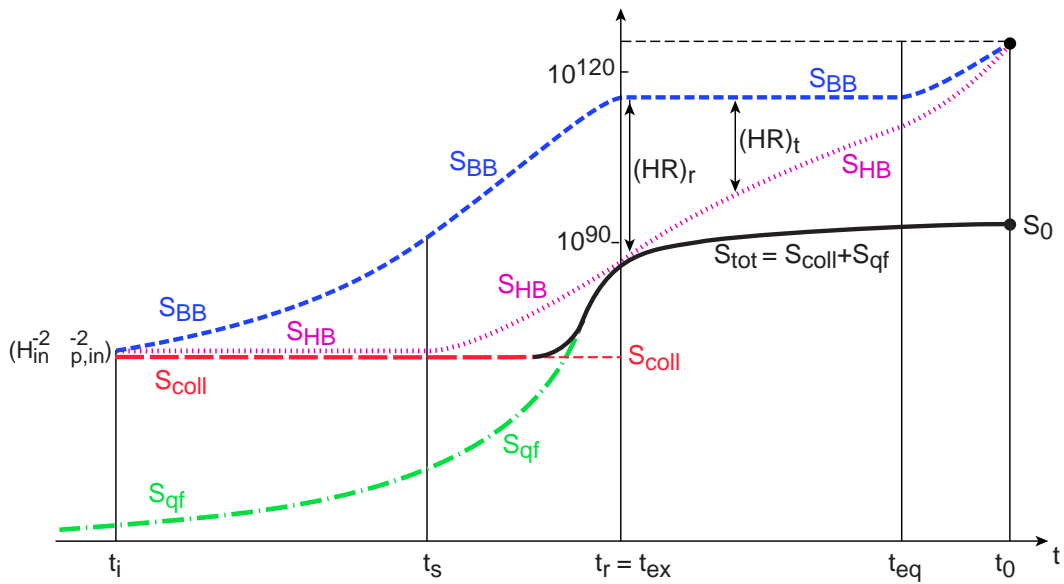,width=14cm}
 \caption[]{At the beginning of the DDI era ($t=t_i$)  the entropy of the just-formed
 black hole, $S_{coll}$, coincides with both
the BEB, $S_{BB}$, and the HEB, $S_{HB}$,
 while the entropy in quantum fluctuations, $S_{qf}$, is completely negligible.
At the beginning of the string phase, $t=t_s$, both  $S_{coll}$ and $S_{HB}$ still have their
common initial value. $S_{BB}$ and $S_{qf}$ have grown considerably, but the latter is still negligible
if the string coupling is still small at $t=t_s$. During the string phase, $S_{qf}$ catches up
with $S_{coll}$ first, and with $S_{HB}$ later, i.e. when the energy
 in the quantum fluctuations becomes critical and
 exit to radiation is expected ($t=t_r$).  Finally, during the radiation and matter-dominated phases,
$S_{HB}$  grows towards $S_{BB}$, while our own entropy $S_{tot}$ lags far behind
 and  increases only slowly as the result of dissipative phenomena and growth of inhomogeneities.}
\end{figure}

It is helpful to follow the evolution of various
entropies with the help of Fig.1.
At time $t=t_i$, corresponding to the first appearance of a horizon, we can use the
Bekenstein--Hawking formula to argue that \begin{equation}
 S_{coll} \sim (R_{in}/l_{P,in})^2 \sim (H_{in} l_{P,in})^{-2} = S_{BB} = S_{HB}  \; ,
\label{initialS}
\end{equation}
where we have used the fact \cite{BDV} that the initial size of the black-hole horizon determines
also the initial value of the Hubble parameter.
Thus, at the onset of collapse/inflation, there is
no hierarchy between the two bounds and entropy is as large as allowed by them.
Furthermore, since the collapsing region is large in string (and a fortiori
in Planck) units, eq. (\ref{initialS}) corresponds to a large number.
Incidentally,  this number is also close to the number of quanta needed for the collapse to occur \cite{BDV}.  We have also assumed
the initial quantum state to
be the ground state. Because of the small initial coupling
and curvature, quantum
fluctuations around it are very small \cite{GPV}, initially,
and contribute a negligible amount $S_{qf}$ to the total entropy.

After a short transient phase, dilaton-driven inflation (DDI) should follow and last until $t_s$,
the time at which a string-scale curvature $O(M_s)$ is reached. We expect the process not
 to generate further entropy (unless more energy flows into the black hole,
 but this would only increase its total comoving volume), but what happens
to the two bounds? This is the crucial observation: while the HEB also stays constant,
the BEB grows, causing a large discrepancy between the two at the end of the DDI phase.
In order to show this, let us recall one of the equations of string cosmology \cite{PBB},
the conservation law:
 \begin{equation}
\partial_t \left(e^{-\phi} \sqrt{g} H \right) =
\partial_t \left((\sqrt{g} H^3)~~( e^{-\phi} H^{-2}) \right) = \partial_t \left( n_H S_H \right) = 0~.
\label{conservation}
\end{equation}
 Comparing with (\ref{HB}), we
recognize that (\ref{conservation}) simply expresses the time independence
of the HEB during the DDI phase. While at the
beginning of the DDI phase $n_H =1$, and the
whole entropy is in a single Hubble volume, as DDI proceeds the same total amount of entropy
becomes equally shared among very many Hubble volumes until, eventually, each one of them
contributes a relatively small number. By contrast, it is easy to see that the BEB is increasing
fast during the DDI phase since, using (\ref{conservation}),
\begin{equation}
S_{BB} \sim MR \sim \rho R^4 \sim H^2  e^{-\phi} \sqrt{g} R = \rm{const.} \times (HR) \; ,
\label{BBgrowth}
\end{equation}
and both $R$ and $H$ grow during DDI.
Also, having assumed that the string coupling is still small at the end of DDI,
we can easily argue that
the entropy in quantum fluctuations remains at a negligible level during that phase.

Something interesting happens if we now consider the string phase \footnote{We concentrate
here on the standard PBB scenario \cite{PBB} and not on its variant \cite{MR} in which the DDI phase
flows directly into a low-energy M-theory phase.}, characterized by a constant
$H$ and $\dot{\phi}$. It is easy to find that, if $\dot{\phi} > 3H$, the HEB starts to decrease
while for $\dot{\phi} < 3H$ it increases. Clearly, the first alternative leads
to a contradiction with the HEB, since $S_{coll}$ cannot decrease. We are thus led to the
amusing result that the HEB demands $\dot{\bar{\phi}} \equiv \dot{\phi} - 3H \le 0$ during the string phase as opposed to the $\dot{\bar{\phi}} > 0$ condition that characterizes the DDI phase. Thus, the HEB implies
 a ``branch change" occurring between the DDI and the string phase, a well known necessary condition for achieving a graceful exit \cite{exit}. The condition $\dot{\bar{\phi}} < 0$ for the string phase also follows from  (apparently independent) arguments based on the study of late-time attractors \cite{MGV,BM}.

When will the final exit to the FRW phase occur? It has been assumed \cite{BBB} that it does when the energy
 in the quantum fluctuations
becomes critical, i.e. when

\begin{equation}
\rho_{qf}  \sim N_{eff} ~ H^4_{max} = e^{-\phi_{exit}} M_s^2 ~ H^2_{max} ~,
\label{BBB}
\end{equation}
where $N_{eff}$ is the effective number of
particle species produced. Taking $H_{max} \sim  M_s$,  fixes the value of the dilaton at exit,
 $e^{\phi_{exit}} \sim 1/N_{eff}$.
Using known results
on entropy production due to the cosmological squeezing of vacuum fluctuations \cite{entropy},
we find:
\begin{equation}
S_{qf}({\rm ex})  \sim N_{eff} ~ H^3_{max} V \sim e^{-\phi_{exit}} M_s^3 ~  V \sim
\left(l_s^2/l_P^{2}\right)_{exit}  ~ V l_s^{-3} \sim  S_{HB}({\rm ex}) ~, \label{qfentropy}
\end{equation}
i.e.  saturation of the HEB by $S_{qf}$. Unless exit occurs at this point, the HEB will be violated at later times.

We thus arrive, generically,
at the situation shown in Fig.1. At $t=t_{ex} \equiv t_r$, the entropy in the quantum fluctuations
 has catched up with (and possibly overcome) that of the classical collapse and has become equal
to the HEB,  $S_{HB} \sim (HR)^3 \sim 10^{90}$. By then, $S_{BB}$
 is a factor $HR$ larger, which is precisely the factor
$10^{30}$ that we are running after. From there on, the story is simple: our entropy remains, to date, roughly constant and around
$10^{90}$, while $S_{HB}$ keeps increasing --with somewhat different rates-- during the radiation
 and the matter-dominated
epochs. $S_{BB}$  always remains a factor $HR$ above $S_{HB}$, but this factor, originally huge,
shrinks to unity today, by definition.

In conclusion, the entropy and arrow-of-time problems are neatly solved, in PBB cosmology,
by the identification of our observable Universe with (part of) the interior of an original
 black hole. As such, its initial entropy saturates both the HEB and the BEB and is large because of the assumed large size (in string or Planck units) of the initial black hole. From there on, there is a natural mechanism to provide
saturation of the HEB at the beginning of the radiation-dominated phase, i.e. when the BEB lies
some thirty orders of magnitude higher. This is precisely what is needed to account for the
initial entropy of our Universe, and to unambiguously identify its time arrow.

We do not wish to conceal the fact that our choice of the initial
size of the collapsing/inflating region can be objected to, along the lines of Refs. \cite{TW},
 as representing a huge amount of fine-tuning. Our answer to this objection  has  been expressed elsewhere \cite{GV}: the classical collapse/inflation process is a scale-free problem in
General Relativity; as such, it  should
 lead to a flattish distribution of horizon sizes, extending from the string length to very large scales, including those appropriate for giving birth to our Universe.
No other dimensionless ratio is tuned to a particularly large or small
value as evidentiated in Fig.1 by the three upper curves all originating from the same point at $t=t_i$.
Finally, we wish to stress again that the entropy considerations discussed in this note appear to provide new general arguments  supporting previous conjectures on the way pre-big bang
inflation should make a  graceful exit into standard, post-big bang FRW cosmology.
\vspace{3mm}

Useful discussions with M. Bowick,  R. Brustein, M. Gasperini, A. Ghosh,
F. Larsen, R. Madden and E. Martinec are gratefully acknowledged.

\vspace{3mm}
Note added: After completion of this work I became aware of 
a very recent paper \cite{Easther}
which reaches  similar conclusions on the role of $H^{-1}$ in cosmological entropy bounds.

\end{document}